\documentclass[a4paper,11pt]{article}

\usepackage{multicol}
\usepackage{graphicx}
\usepackage{setspace}
\usepackage[
   pdftitle={$\rm 5s$ correlation confinement resonances in Xe-endo-fullerenes}, %%% <- your title
   pdfauthor={V. K. Dolmatov and D. A. Keating},         %%% <- your author list
   pdfkeywords={photoionization, correlation, endoheadral atoms},   %%% <- your keywords
   colorlinks,
   linkcolor=blue,
   urlcolor=blue,
   citecolor=blue,
   a4paper=true,
dvipdfm           % Uncomment if using dvipdfm to convert dvi into pdf
   ]{hyperref}

\newcommand{\abstracttitle}[1]{
 \begin{center}{\Large {\bf #1}}\end{center}
}
\newcommand{\authors}[1]{
 \vspace*{-0.3cm}
 \begin{center} {\bf #1} \end{center}
 \vspace*{-0.3cm}
}
\newcommand{\addresses}[1]{
 \begin{center} {\small #1} \end{center}
}
\newcommand{\synopsis}[1]{
 \begin{center}
 \setstretch{0.75}
 \begin{minipage}[t]{16cm}
   {\footnotesize {\bf Synopsis} #1 }
 \end{minipage}
 \setstretch{1.0}
 \end{center}
}
\newcommand{\abstracttext}[1]{
 \vspace*{-0.3cm}
 \columnsep0.75cm
 \begin{multicols}{2} #1 \end{multicols}
}
\newcommand{\picturelandscape}[2]{
 \vspace*{0.5cm}
 \centerline{
  \includegraphics*[width=7.8cm,angle=#1]{#2}
 }
}
\newcommand{\capt}[2]{
 \vspace*{-0.3cm}
 \begin{center}
 \begin{minipage}[t]{7.8cm} {\small {\bf Figure~#1}.~#2} \end{minipage}
 \end{center}
 \vspace*{0.3cm}
}

\newcommand{\writeto}[1]{
 \hspace*{-2.5mm} \footnote{E-mail: \href{mailto:#1}{#1}}\hspace*{-1.5mm}
}

\begin{document}

\abstracttitle{$\rm 5s$ correlation confinement resonances in Xe-endo-fullerenes
% Instructions for preparation of abstracts for ICPEAC 2011 \\
% second line of title if needed
}

\authors{
V. K. Dolmatov\writeto{vkdolmatov@una.edu}
 and D. A. Keating
% and so on ...
}

\addresses{Department of Physics and Earth Science, University of North Alabama, Florence, AL 35632, U.S.A.}

\synopsis{Spectacular trends in the modification of the Xe $\rm 5s$ photoionization via interchannel coupling with confinement resonances emerging in the Xe $\rm 4d$ giant resonance
upon photoionization of the  Xe@C$_{60}$, Xe@C$_{240}$ and Xe@C$_{60}$@C$_{240}$ endo-fullerenes are theoretically unraveled and interpreted.}

\abstracttext{Resonances, termed \textit{confinement resonances} (CR's), emerging in photoionization spectra of  atoms $A$ confined inside the
empty spaces of endo-fullerenes - $A$@C$_{n}$, $A$@C$_{n}$@C$_{m}$, \textit{etc} - have attracted much attention of investigators in recent years \cite{AQC09} (and references therein).
CR's occur in endo-fullerene photoionization spectra due to the interference of the photoelectron waves
emerging directly from the confined atom $A$, and those scattered off the endo-fullerene confining cage(s). Their existence has recently been confirmed
experimentally \cite{Phaneuf2010}. \textit{Correlation confinement resonances} (CCR's) differ from ordinary CR's in that
they occur in the photoionization of an \textit{outer} subshell of the confined atom $A$ due to the interference of transitions from the outer subshell  with ordinary CR's emerging in
\textit{inner} shell photoionization transitions, via interchannel coupling~\cite{AQC09,CCR}. They were originally
theoretically discovered in the Xe $\rm 5s$ photoionization of Xe@C$_{60}$ \cite{CCR}. The CCR phenomenon thus introduces a novel class of resonances that can exist neither without
a confinement nor electron correlation, thereby attracting much interest to their study. 

Ordinary CR's in the Xe $\rm 4d$ giant photoionization resonance
undergo dramatic changes along the path from Xe@C$_{60}$ to Xe@C$_{240}$ to Xe@C$_{60}$@C$_{240}$ \cite{ICPEAC}. With the impetus of results of
Ref.~\cite{ICPEAC}, we study how ordinary CR's in the Xe $\rm 4d$ photoionization affect the Xe $\rm 5s$ photoionoization in various Xe-endo-fullerenes.
It is the aim of this paper to report on spectacular CCR's emerging in the Xe $\rm 5s$ photoionization cross section due to interchannel coupling of the $\rm 5s$ $\rightarrow$ $\rm p$
transition with transitions from the Xe inner $\rm 4d$ subshell, and how these CCR's change along the path Xe@C$_{60}$ $\rightarrow$ Xe@C$_{240}$ $\rightarrow$ Xe@C$_{60}$@C$_{240}$.

In performed
calculations, fullerene cages were approximated by square-well potentials of certain inner radii $r_{0}$, widths $\Delta$
and depths $U_{0}$ \cite{AQC09,CCR}. The Xe atom was placed at the center of fullerenes, both single-cage and nested fullerenes. The updated values of $\Delta \approx 1.25$, $r_{0} \approx 6.01$ and $U_{0} \approx - 0.422$\ {\em au} \cite{ICPEAC} for C$_{60}$
were used in calculations. This is because they were shown~\cite{ICPEAC} to provide a closer agreement between experiment theory.
As for C$_{240}$, $\Delta \approx 1.25$, $r_{0} \approx 12.875$ and $U_{0} \approx - 0.52$\ {\em au} \cite{ICPEAC}.  Interchannel coupling was accounted for in the
framework of the random phase approximation with exchange~\cite{Amusia}.
The calculated data are depicted in figure $1$ where spectacular changes in the Xe $\rm 5s$ photoionization along the path
Xe@C$_{60}$ $\rightarrow$ Xe@C$_{240}$ $\rightarrow$ Xe@C$_{60}$@C$_{240}$ are seen. A  detailed discussion will be offered at the spot.

 \picturelandscape{0}{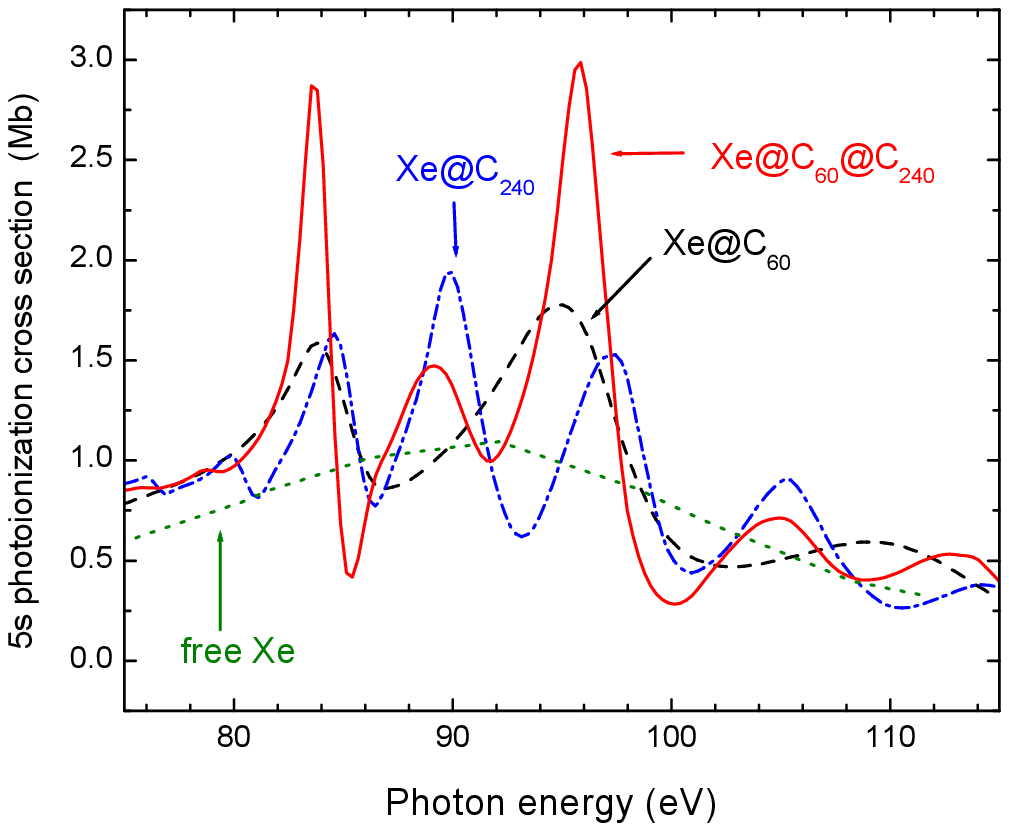}
%\picturelandscape{0}{Min_Max_Average_Belfast.pdf} % Use this for pdflatex
\capt{1}{Correlation confinement resonances in the Xe ${\rm 5s}$ photoionization cross section of Xe@C$_{60}$,
 Xe@C$_{240}$ and Xe@C$_{60}$@C$_{240}$, as marked. They ``mirror'' CR's in the Xe $\rm 4d$ photoionization
 cross sections (not shown) of the endo-fullerenes in questions \cite{ICPEAC}, due to interchannel coupling.}

This work was supported by the NSF Grant No.\ PHY-0969386
and a UNA CoA\&S  grant.

}  % end of the body
\end{document}